\documentstyle[twocolumn,aps,epsfig,times,latexsym,amssymb]{revtex}

\title{Design of Strongly Modulating Pulses to Implement \\
Precise Effective Hamiltonians for Quantum Information Processing}
\author{Evan M. Fortunato$^{\dag}$,  Marco A. Pravia $^{\dag}$,  Nicolas Boulant$^{\dag}$,\\
Grum Teklemariam$^{\ddag}$, Timothy F. Havel$^{\dag}$, David G. Cory$^{\dag}$}
\address{$^{\dag}$Department of Nuclear Engineering, Massachusetts Institute of Technology,
Cambridge, MA 02139\\
$^{\ddag}$Department of Physics, Massachusetts Institute of Technology, Cambridge, MA 02139}

\begin{document}

\def\beqn{\begin{equation}}
\def\eeqn{\end{equation}}
\newcommand{\dt}{\! \cdot \!}
\newcommand{\etal}{{\em et al. }}
\newcommand{\ket}[1]{$\vert${#1}$\rangle$}
\newcommand{\mket}[1]{\vert{#1}\rangle}
\newcommand{\tfrac}[2]{{\textstyle\frac{#1}{#2}}}

\maketitle

\begin{abstract}
We describe a method for improving coherent control through the use
of detailed knowledge of the system's Hamiltonian.  Precise unitary 
transformations were obtained by strongly modulating the system's dynamics to 
average out unwanted evolution.  With the aid of numerical search methods, pulsed 
irradiation schemes are obtained that perform accurate, arbitrary, selective 
gates on multi-qubit systems.  Compared to low power selective pulses, which 
cannot average out all unwanted evolution, these pulses are substantially 
shorter in time, thereby reducing the effects of relaxation.  Liquid-state NMR 
techniques on homonuclear spin systems are used to demonstrate the accuracy 
of these gates both in simulation and experiment.  Simulations of the coherent 
evolution of a 3-qubit system show that the control sequences faithfully 
implement the unitary operations, typically yielding gate fidelities on the 
order of $0.999$ and, for some sequences, up to $0.9997$.  The experimentally 
determined density matrices resulting from the application of different 
control sequences on a 3-spin system have overlaps of up to $0.99$ with the 
expected states, confirming the quality of the experimental implementation.  
\end{abstract}

\section{Introduction}
The past decade has seen a substantial interest in improving coherent 
quantum control.  Coherent control had origins in both nuclear magnetic 
resonance (NMR)~\cite{Rabi,Waugh} and optical spectroscopy
~\cite{HaughtControl,WarrenControl}.  For an overview of advances
in both fields, the reader is referred to~\cite{WarrenReview}.
Since coherent control's inception, many different 
techniques have been used both to improve selectivity and to reduce the 
duration of control pulses.  For spin systems, the Fourier transform 
has been used to approximate the excitation profile in the limit of low 
power and no spin-spin couplings~\cite{Sodickson} and more complete analytic 
solutions have been developed to aid in general pulse design and analysis
~\cite{Counsell,Conolly,Tycko,Morris,Sorensen,Emsley}.  Alternatively, very 
sophisticated shaped pulses have been designed using a variety of 
computer-aided methods~\cite{Green,Ewing} or feedback from system 
observables~\cite{Schiano}.  Equivalent analytic theories~\cite{Peirce,Zewail}, 
computer-aided methods~\cite{Dahleh} and feedback techniques
~\cite{Assion,Pearson} have also been developed by the optics community.
Similar techniques are also used in other fields such as the control of
trapped ions~\cite{Molmer,Wineland}.

The development of liquid-state NMR systems as prototype quantum
information processors~\cite{CoryPP,ChuangPP} has enabled experimental
demonstration of quantum algorithms~\cite{DJ,Grover,QFT}, quantum
error correction~\cite{3BitQEC,5BitQEC}, and quantum
simulations~\cite{QSim}.  These experiments built upon
well-established spectroscopic techniques developed over the past four
decades, such as using low-power (soft) shaped radio-frequency (RF)
pulses to obtain selective operations.  However, the selective pulses
employed to date have the disadvantage that low power implies long
duration. This not only introduces errors due to relaxation, or
decoherence, but also allows significant evolution under the action of
the internal Hamiltonian.  In the past, this evolution was rarely of
concern because there was little importance placed on implementing a
particular operation.  For example, in spectroscopy there are entire
classes of propagators that selectively excite a single spin from its
equilibrium state, but for applications such as quantum computing the
transformation must act as expected for all input states. A second
problem with soft-pulse techniques is that selective pulses
simultaneously applied to different spins interfere with each other,
thus causing significant deviation from the desired
action~\cite{Bloch-Siegart}.  To address some of these problems,
several groups precalculate these errors and incorporate corrections
into their analysis and pulse design ~\cite{ChuangPulses,KnillPulses}.
However, not all errors can be corrected using these
techniques~\cite{PulseProblem}, and it would be preferable to average
out unwanted evolution by the use of strong control fields, so that no
additional corrections are required.

In this paper, we present a procedure for finding high-power pulses that
strongly modulate the system's dynamics to produce precisely a desired 
spin-selective unitary propagator.  These operations, or gates, allow arbitrary 
rotations of each spin around independent single-spin axes, while refocusing the 
internal evolution. They are ``self-contained,'' in the sense that they can be 
placed back to back in longer sequences without requiring additional computational 
resources or post-experiment corrections. By using high-power, pulse durations 
are decreased by almost an order of magnitude, thereby significantly reducing 
the effects of relaxation. Finally, the use of strong modulation also opens 
the possibility of incorporating robustness against slowly varying or time
independent incoherent errors such as those caused by static field or RF 
inhomogeneities~\cite{Hahn,CP,MG}. Our control methods are the first to combine 
all of these features.

The pulses presented here have been applied in recent Quantum Information 
Processing (QIP) experiments to demonstrate algorithms~\cite{QFT}, study 
notions of measurement~\cite{Eraser}, and test new methods for noise 
control~\cite{NS}.  They promise to be increasingly useful in future NMR QIP 
experiments, where larger numbers of qubits will necessitate increasing the 
number of homonuclear spins.  In addition, these methods can be adapted to 
develop improved pulses for selective spectroscopy~\cite{Spect} and 
imaging~\cite{image}. Finally, although presented within the context of NMR, 
these methods are applicable to any system where the total Hamiltonian is 
well known and the external degrees of freedom allow for universal 
control, both requirements of any quantum information processor.

\section{Control of Quantum Information}

In the standard model of quantum computing, an algorithm can be expressed 
as a series of unitary operations that maps a set of input states to a 
particular set of output states.  The physical implementation of an 
algorithm requires the use of a quantum system with a Hamiltonian that 
contains a sufficient set of externally controlled parameters to allow for 
the generation of a universal set of gates~\cite{Universality}.  The task 
of control is to find a time-dependent sequence of values for these control 
parameters that modulates the system's dynamics in order to generate a 
particular gate to the required precision.

Given a control sequence, solving for the effective Hamiltonian is
straightforward.  Unfortunately, going the other way is much more
difficult.  That is, finding an RF waveform that produces a propagator
with desired properties is an inverse problem.  Traditionally,
analytic techniques, such as average Hamiltonian theory~\cite{Waugh},
have been used to determine an appropriate control sequence.  With
modern computer resources, numerical methods provide a more efficient
and accurate solution to this problem.  

\subsection{Gate Fidelity as a Metric for Control}
A metric of a gate's performance should describe the quality of a
general transformation, including the possibility of non-unitary
evolution.  Unfortunately, such information is not conveniently 
accessible by experiment, so we choose a metric comprised only of 
sets of state measurements.  For an input state, $\rho_{in}$, 
the ideal transformation maps the system to a theoretical output state,
$\rho_{th}$, {\it i.e.}, 
\beqn 
\label{rho_th}
\rho_{in} \longrightarrow \rho_{th}.  
\eeqn 
On the other hand, a simulated or experimentally implemented control 
sequence will produce a different output state, 
$\rho_{out}$, {\it i.e.}, 
\beqn 
\label{rho_out}
\rho_{in} \longrightarrow \rho_{out}.  
\eeqn 
Noting that $\rho$ is Hermitian, the projection between these two states, 
defined as 
\beqn 
P(\rho_{th}, \rho_{out}) = \frac{trace(\rho_{th}~\rho_{out})}
	{\sqrt{trace(\rho_{th}^2) trace(\rho_{out}^2)}},
\eeqn 
quantifies how similar in 'direction' the two states are.  This 
metric is analogous to the dot product between two vectors, varying 
from $-1$ for anti-parallel states to $1$ for identical states.  A 
value of zero indicates orthogonal density matrices.  In order to 
account for non-unitary evolution, a second term multiplies the 
projection yielding the attenuated correlation, namely, 
\begin{eqnarray}
C(\rho_{th}, \rho_{out}) = P(\rho_{th}, \rho_{out}) 
	\sqrt{\frac{trace(\rho_{out}^2)}{trace(\rho_{in}^2)}}\\
	= \frac{trace(\rho_{th}~\rho_{out})}
	{\sqrt{trace(\rho_{th}^2) trace(\rho_{in}^2)}}.
\end{eqnarray}
We define the gate fidelity, F, of a transformation as 
\beqn
\label{GateFidelity} 
F = \overline{C(\rho_{th},\rho_{out})}, 
\eeqn 
where $\overline{C}$ represents the average attenuated correlation 
over an orthonormal set of input density operators ({\it i.e.}, 
$Trace[\rho_j\rho_k]=\delta_{jk}$) that span the Hilbert
space.  It should be noted that $F$ is maximized (with a value of one)
when the implemented and ideal transformations are the same, and is
insensitive to differences in the global phase between the ideal and
implemented transformation.  

We can derive a useful alternate form for the gate fidelity
in terms of the actual and theoretical transformations instead of the
input-output state relations.  This form is both easier to compute and 
has intuitive appeal in that knowledge of the transformation can be
directly translated to gate fidelities.  First we assume that our
ideal transformation is unitary, and the implemented transformation is 
a completely positive, trace-preserving linear map~\cite{Schumacher}.  
In other words, the implemented transformation takes normalized density 
operators to normalized density operators and if the system starts as 
subsystem of an entangled system, then the full system's density operator 
also maps in a reasonable way.  Under these assumptions, Eq. (\ref{rho_th}) 
and (\ref{rho_out}) are explicitly given by 
\beqn
\rho_{th}=U_{th} \rho_{in} U_{th}^{\dagger},
\eeqn
and
\beqn
\rho_{out}=\sum_{\mu} A_{\mu} \rho_{in} A_{\mu}^{\dagger}, 
\eeqn
where the $A_{\mu}$ satisfy
\beqn
\sum_{\mu}A_{\mu}^{\dagger}A_{\mu}=\openone.
\eeqn
We now show that the gate fidelity reduces to 
\beqn 
\label{ReducedF}
F = \sum_{\mu} |Trace(U_{th}^{\dagger}A_{\mu})/N|^2,
\eeqn
where N is the dimension of the Hilbert space. 
Using the normalized Pauli basis, $\sigma_j$,  as the orthonormal 
input density operators and the cyclic properties of the trace,
Eq. (\ref{GateFidelity}) becomes
\begin{eqnarray}
F=\sum_{j=1}^{N^2} Tr[(U_{th} \sigma_j U_{th}^{\dagger}) 
	(\sum_{\mu} A_{\mu} \sigma_j A_{\mu}^{\dagger})] / N^2 \\
\label{FullFidelityForm}
=\sum_{j=1}^{N^2} Tr[\sigma_j \sum_{\mu} U_{th}^{\dagger} A_{\mu} 
	\sigma_j A_{\mu}^{\dagger} U_{th}] / N^2.
\end{eqnarray}
Expanding the product of $U_{th}^{\dagger}A_{\mu}$ 
in terms of the orthonormal Pauli basis 
($U_{th}^{\dagger}A_{\mu} = \sum_{k} B_{\mu}^k\sigma_k$), yields
\begin{eqnarray}
=\sum_{j \mu} Tr[\sigma_j (\sum_{k} B_{\mu}^k \sigma_k) 
		 \sigma_j (\sum_{m} {B_{\mu}^m}^* \sigma_m)] /N^2 \\
\label{FC}
=\sum_{j \mu k m} B_{\mu}^k {B_{\mu}^m}^* 
	Tr[\sigma_j \sigma_k \sigma_j \sigma_m] /N^2.
\end{eqnarray}
Because the $\sigma$ basis is orthogonal, only terms where $k=m$ contribute.
Therefore, Eq. (\ref{FC}) reduces to
\beqn
\label{FC2}
F=\sum_{j \mu k} |B_{\mu}^k|^2 Tr[\sigma_j \sigma_k \sigma_j \sigma_k]/N^2.
\eeqn
If $\sigma_k$ is not proportional to identity, it will anti-commute 
with exactly half the $\sigma_j$ terms in the sum, while commuting with
the other half.  Therefore, two sets of terms cancel and have 
no contribution to F.  
Defining $\sigma_1$ to be the element that is proportional 
to identity, Eq. (\ref{FC2}) further reduces to
\begin{eqnarray}
F= \sum_{j \mu} |B_{\mu}^1|^2 Tr[\sigma_j\sigma_j]/N^3 \\
 = \sum_{j \mu} |B_{\mu}^1|^2/N^3 = \sum_{\mu} |B_{\mu}^1|^2/N.
\end{eqnarray}
This is clearly equal to Eq.~\ref{ReducedF}.
Thus, the gate fidelity corresponds to how well the actual transformation 
reverses the action of $U_{th}^{\dagger}$.  In this form it is obvious that the 
gate Fidelity is independent of which orthonormal basis of input states are used 
as $\rho_{in}$.

\subsection{NMR as an Example System}
As an example, liquid-state NMR is used to demonstrate how to find control 
sequences to implement particular gates.  In NMR, spins in a large static 
magnetic field (in our case, 9.7 T) are controlled via external RF pulses.  The internal 
spin Hamiltonian is composed of both Zeeman interactions with the applied field 
modified by electron screening (chemical shift) and scalar couplings 
with other spins.  Together these provide the QIP requirements of 
addressability and conditional logic respectively.  
In terms of spin operators, the internal Hamiltonian is
\begin{equation}
H_{int}=\sum_{k=1}^n \omega_k I^k_z + 2\pi\sum_{j>k}^n\sum_{k=1}^n J_{kj} \\
I^{k}{\cdot}I^{j},
\end{equation}
where $\omega_k$ represent the chemical shifts of the spins, $J_{kj}$ the 
coupling constant between spins $k$ and $j$, and $n$ is the number of spins.
The external Hamiltonian describing the coupling between the spins and an 
oscillating RF field generated by a single transmitter is
\begin{equation}
H_{ext}(\omega_{RF}, \phi, \omega, t)=\sum_{k=1}^n e^{-i(\omega_{RF} t+\phi)I^k_z }\\ 
	(-\omega I^k_x) e^{i(\omega_{RF}t+\phi)I^k_z } ,
\end{equation}
where $\omega_{RF}$ is the transmitter's angular frequency, $\phi$ the initial phase, 
and $\omega$ the power~\cite{PowerNote}.  Of course, additional species can be added 
by including appropriate terms in $H_{int}$ and an additional $H_{ext}$ for each 
additional RF field.

Using this knowledge of the internal Hamiltonian and the form of the external 
Hamiltonian, the parameter values that generate the desired gate must be 
determined.  Here, a quality factor $Q=1-\sqrt{F}$ is minimized by searching 
through the mathematical parameter space using the Nelder-Mead Simplex 
algorithm~\cite{Simplex}.  While this function has many local minima, the 
Simplex algorithm often succeeds in finding satisfactory solutions.  Our goal 
is to show that sufficient, implementable control sequences can be found.  
Finding the optimal solution is much more challenging and based on our
system and control parameter values, is not expected to improve pulse 
performance significantly.  We have parameterized the control sequence as a cascade 
of RF pulses with fixed power, transmitter frequency, initial phase, and pulse 
duration ($\tau$).  As will be seen, this is a particularly convenient and completely 
general parameterization, but we make no claims that it is the only, nor necessarily 
the best choice.  If the RF power is constant over the duration of a pulse, 
{\it i.e.}, the pulse's amplitude is square, the total Hamiltonian 
$H_{tot}=H_{int}+H_{ext}$ can be made time independent by transforming into the 
frame that rotates at the frequency of the transmitter.  This allows the 
Liouville-von Neumann equation of motion to be solved by a single diagonalization.  
Initially, the starting density matrix is the same in both frames 
($\tilde{\rho}(0) = \rho(0)$), so that at the end of the pulse, the density 
matrix in the new frame is given by
\begin{equation}
\tilde{\rho}(\tau) =e^{-iH_{eff} \tau} \rho(0) e^{iH_{eff} \tau},
\end{equation}
where $H_{eff}$ is the effective Hamiltonian in the new frame~\cite{ErnstBook}.  
Transforming this density matrix back to the original rotating frame gives
\begin{equation}
\rho = U_z(\tau)^{-1} e^{-iH_{eff} \tau} \rho(0) e^{iH_{eff} \tau} U_z(\tau),
\end{equation}
where 
\begin{equation}
U_z(\tau)=(e^{i\omega_{RF}\sum_{k=1}^n I^k_z \tau}).
\end{equation}
Therefore, in the original rotating frame, the transformation is given by 
\begin{equation}
U_{period}(\tau)= U^{-1}_z(\tau)e^{-iH_{eff} \tau}.
\end{equation}
Because the evolution under the whole sequence is given in the
original rotating frame, no additional resources are required to
concatenate pulses, nor is any mathematical correction required at 
the end of an experiment.

Cascading these periods yields the net transformation
\begin{equation}
U_{net}=\prod_{m=1}^N U^{-1}_z(\tau_m) \\
	e^{-iH_{eff}^m(\omega^m,\omega_{RF}^m,\phi^m) \tau_m}
\label{UNET}
\end{equation}
where the index $m$ refers to the $m^{th}$ period, {\it i.e.}, to the
$m^{th}$ square pulse, with a corresponding set of 4 parameters. In other 
words, $N$ constant amplitude pulse periods, each with a different transmitter 
frequency and initial phase, are applied in series. Clearly, a single period 
is not sufficient to generate an arbitrary transformation; therefore the number 
of periods is increased until a suitable net transformation is found.  Using 
desktop computing resources, this yields convergence times for three- and 
four-spin systems that are typically seconds to minutes.

In addition to the desired propagator, $U_{ideal}$, an initial set of starting 
parameters for the pulse shape is required.  While this initial guess must be 
reasonable ({\it i.e.}, in the vicinity of the solution), many different starting 
points typically converge to equally deep minima.  We have observed that the 
number of acceptable solutions for this parameterization is very large, allowing 
experimental implementation issues to be considered.  For example, experimental 
limitations do not allow arbitrarily high powers or frequencies to be implemented.  
To keep the algorithm from returning infeasible
solutions, a penalty function that increases as the parameter value moves towards 
infeasible solutions is added to the quality factor.  Penalty functions are 
also used to guide the algorithm towards more favorable pulse solutions.  In 
our case, penalties are placed on high powers, large frequencies, and 
negative- or long-time periods.  

\section{Simulations}
The methods described above were used to obtain a set of pulses that 
implement each of a set of important single-spin gates.
To study the performance of these gates, propagators for each of the pulses 
were simulated under different conditions.  First, the performance of the 
pulses under idealized experimental conditions is considered.  Second, the 
gate fidelity is simulated as a function of systematic distortions of the
pulse parameters.  From these results, the relative importance of
implementation precision is determined.  Finally, simulations show
that a pulse generates quite different evolutions as a test spin's
resonant frequency is varied over a range of chemical shifts.

\subsection{Ideal Pulse Simulations}
\begin{figure}
		{\centerline{\epsffile{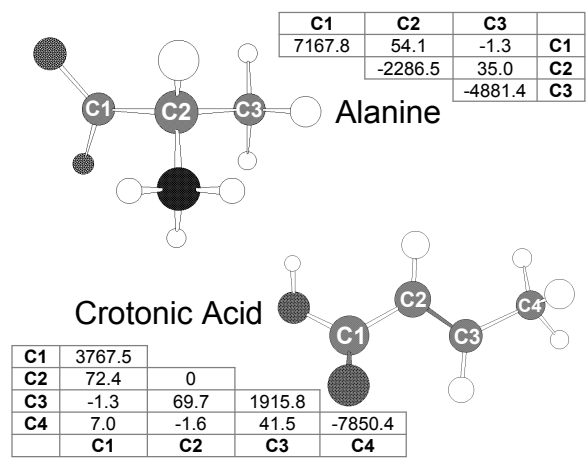}}}
		\vspace{.25cm}
	\caption{Molecular structure and Hamiltonian parameters for both Alanine 
	and Crotonic acid.  The chemical shift of each of the carbon nuclei is 
	given by the corresponding diagonal elements while the coupling strengths 
	are given by the each of the different off-diagonal elements.}
        \label{Molecules}
\end{figure}

\begin{figure}
		{\centerline{\epsffile{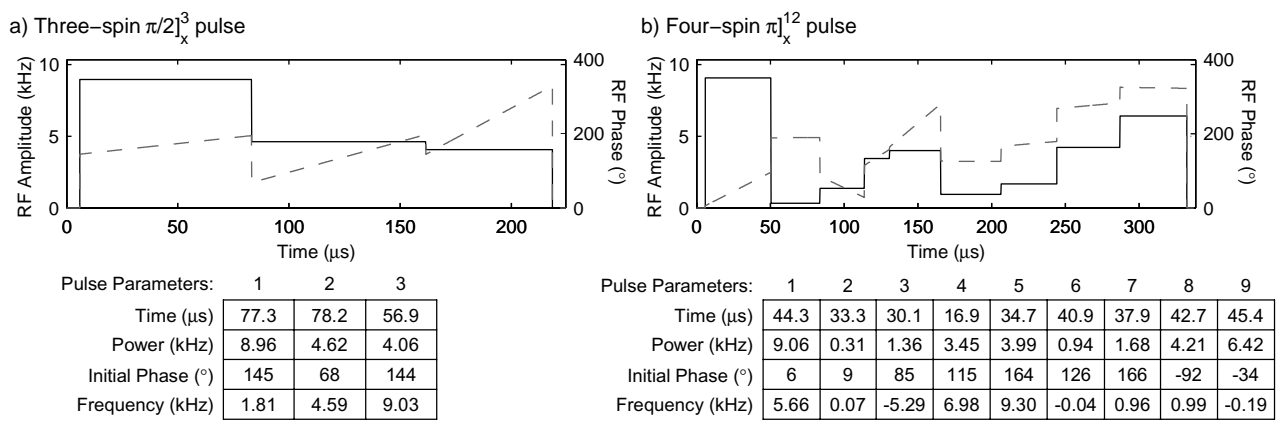}}}
		\vspace{.25cm}
	\caption{The ideal RF waveform for two example pulses.  The solid (dashed) 
	line is the amplitude (phase) of the waveform.  Changes in the transmitter 
	frequency (within a single period) were implemented by a discrete linear 
	phase ramp.  The sharp discontinuities occur at the transitions between 
	periods.  Substantial filtering of these high frequency components (smoothing 
	of the shape) has little effect on the gate fidelity.  In order to experimentally
	implement the pulse, it is converted into a discrete series of amplitudes and 
	phases (order 1K long) by sampling the ideal waveform at a constant rate.  Details 
	of the pulse parameters (as per Eq.~\ref{UNET}) are listed below each waveform.  
	Due to experimental implementation issues, a 6$\mu$s period with zero RF power 
	({\it i.e.,} $H_{ext}=0$) is needed before and after the pulse (see waveform) 
	and must be included to produce the desired propagator.
	}
	\label{pulsesFig}
\end{figure}

Pulses were created for three- ($^{13}C$-labeled Alanine) and four-
($^{13}C$-labeled Crotonic acid) spin homonuclear systems.  The chemical 
shifts and scalar coupling constants for each of these systems are listed 
in Fig.~\ref{Molecules}.  As a representative set, each of the single spin 
$\pi/2$, and nearest-neighbor paired $\pi$ pulses were simulated 
with the relevant characteristics summarized in Table~\ref{PulseChar} and 
example waveforms shown in Fig.~\ref{pulsesFig}.  
The duration of the pulses are on the order of $200 \mu$s for the three-spin 
system and $420 \mu$s for the four-spin system, both significantly shorter 
than those that could be obtained using low-power pulses.  The average 
fidelities for each system are $0.9995$ and $0.995$, demonstrating that, at 
least under ideal conditions, control sequences that implement the desired 
transformation with high fidelity can be found.

The ultimate goal of control in quantum computing is to attain fault-tolerant 
computation.  While it has been proven that perfect control is not 
required~\cite{Fault-Tolerance}, estimates of the precision needed vary from $0.9999$ 
to $0.999999$ depending on the assumptions used.  These simulations predict 
an achievable level of control that approaches the most optimistic estimates 
for fault tolerant computation.  As expected, the pulse duration decreases with 
increasing chemical shifts dispersion (selectivity condition) and, for the case
that $J_{jk} << \left|\omega_k - \omega_j\right|$, the fidelity 
of the sequence decreases with increasing ratio of the couplings (bilinear terms) 
to chemical shift.

\subsection{Variations in the External Hamiltonian}
\begin{figure}
		{\epsfxsize=6.5in\centerline{\epsffile{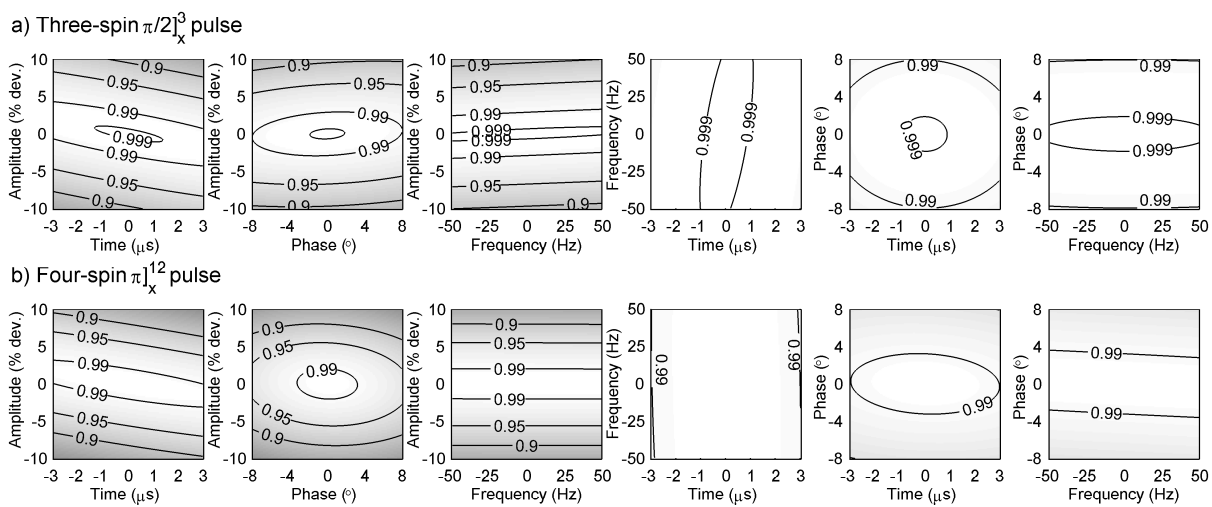}}}
		\vspace{.25cm}
	\caption{Contour plots of the gate fidelity as each pair of 
	parameters is varied over typical experimental errors, demonstrating  
	that both the Alanine and Crotonic acid pulses are most sensitive
	to errors in the RF amplitude.}
	\label{VarExtHamFig}
\end{figure}

The external RF parameters are determined by a minimization procedure,
suggesting that small variations of the external parameters should
have little effect on the quality of the pulses.  To check this
assumption, the gate fidelity was calculated as each of the six pairs
of the four control parameters were varied over a range of errors
typical of an experimental implementation.  As a sample set, one pulse
for each of the two systems is presented here.  The results shown in
Fig.~\ref{VarExtHamFig} demonstrate the natural robustness against
typical variations in the initial phase, frequency, and
duration of each period.  Clearly, the sequence is most sensitive to
power variations.  For the pulses listed in Table~\ref{PulseChar},
if the power's amplitude is changed by $5\%$ the average fidelity falls 
to $0.96 \pm 0.01$ for Alanine pulses and $0.94 \pm 0.04$ for Crotonic 
acid pulses.  For the 25 pulses used in~\cite{NS} the average fidelity at 
$5\%$ amplitude deviation is $0.97 \pm 0.02$.  For $10\%$ deviation, the 
gate fidelity drops to $0.86 \pm 0.03$ for the Alanine pulses and 
$0.81 \pm 0.12$ for the Crotonic acid pulses.  This pulse sensitivity to RF 
amplitude suggests that RF inhomogeneity may be a leading cause of experimental 
errors.  While techniques to select homogeneous regions are 
available~\cite{KnillPulses,CoryRFSelection}, the loss in signal to noise 
is significant, especially if multiple coils are used.  Instead, because these 
errors are incoherent in nature, it is possible to design pulse sequences that 
refocus such homogeneities.  Preliminary results confirm that sequences can
be re-optimized in the presence of significant inhomogeneity.

\subsection{Variations in the Internal Hamiltonian}
\begin{figure}
		{\epsfxsize=3in\centerline{\epsffile{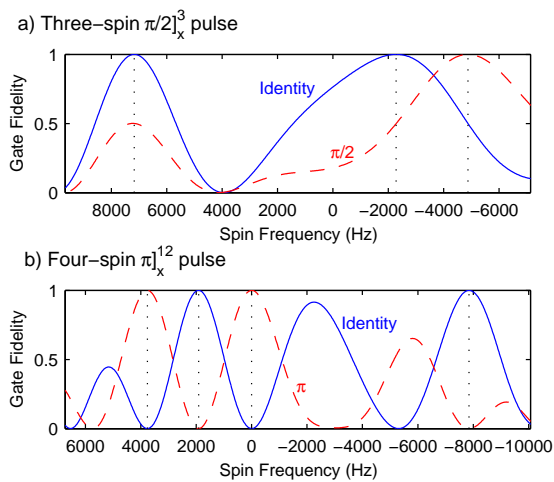}}}
		\vspace{.25cm} 
	\caption{Gate fidelity of two example pulses as the resonance
	frequency of a test spin is varied over a range of 
	chemical shifts.  The solid (dashed) line is calculated with
	identity (desired transformation) as the theoretical transformation.
	The vertical dotted lines denote the actual chemical shifts for
	each spin. 
	As can be seen, the gate only works when the test spin is at the
	appropriate resonance frequency.}
   \label{VarIntHamFig}
\end{figure}

For NMR spectroscopy, the goal is to excite selectively spins in a band of 
frequencies leaving all other possible spins (with unknown precession 
frequencies) along the $z$ axis.  This requires that the propagator 
for spins at any other frequency be, at most a phase change.
With detailed knowledge of the internal Hamiltonian, the effect of the
applied RF field needs only be considered at the resonance frequencies
of the chemical species present in the given molecule.  By relaxing
the requirement that the effective Hamiltonian be zero for all
chemical shifts other than those in the band of excitation, an RF shape
can be found that more efficiently implements the desired gate for the
frequencies of concern yielding high-power yet selective pulses.  To
demonstrate this idea more clearly, the gate fidelities of the two
sample pulses considered in the sub-section B were calculated as a
function of a test spin's resonant frequency.  As can be seen in
Fig.~\ref{VarIntHamFig}, the fidelity is close to unity only near
the resonance frequency for which the pulse was designed to work.
This stresses the necessity of having accurate knowledge of the
system's Hamiltonian.  On the other hand, looking at the region
immediately around the resonance we see the fidelity falls off quite
slowly.  This implies that small variations in the chemical shift do 
not significantly affect the fidelity of the pulses.
For example, in the experiments presented below, the unwanted scalar couplings 
to the hydrogen atoms, which are equivalent to errors in the resonance frequency, 
were automatically refocused by the control pulse.
It should be noted that no constraint was used to require this robustness,
but that it results from the use of strongly modulating pulses.  If
this natural robustness is not sufficient, additional constraints can be
added.  Of course, this robustness also implies that selective pulses will 
be harder to design for spin systems where the spectrum is dense.

\section{Experimental Demonstrations}
In this section we experimentally demonstrate the efficacy of a representative 
set of pulses via both spectra and reconstructed density matrices.  Each waveform 
was discretized by sampling at a constant rate that was faster than the largest 
phase modulation during any of the periods (sample times of order $0.5\mu s$).  
Simulations of the sampled waveform confirmed that the decrease in the gate 
fidelity was acceptably small.  All experimental tests were carried out on the 
three carbons of $^{13}C$-labeled Alanine (see Fig.~\ref{Molecules} for internal 
Hamiltonian parameters) using a Bruker Avance spectrometer.

\begin{figure}
		{\centerline{\epsffile{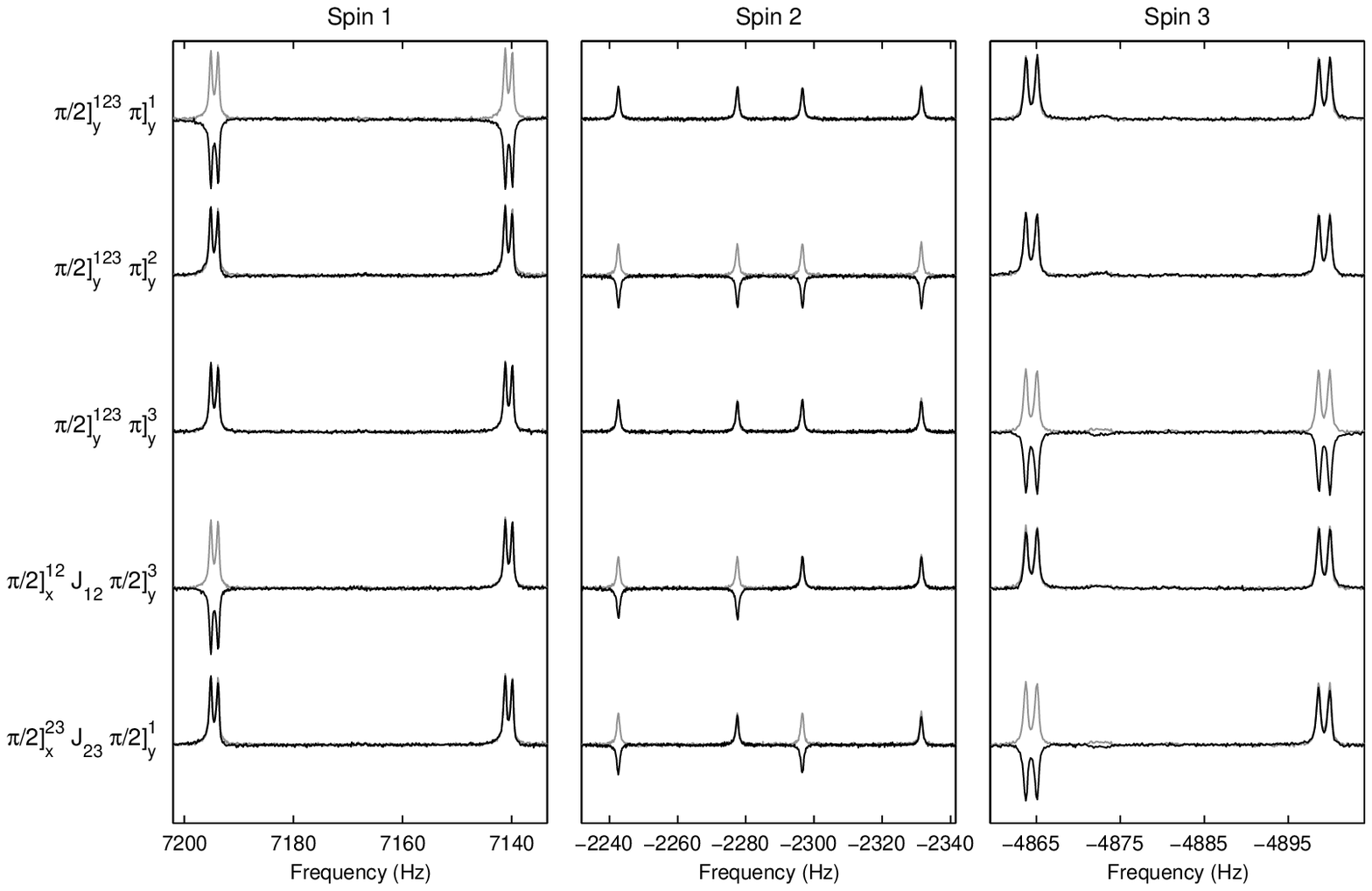}}}
		\vspace{.25cm} 
	\caption{Spectra resulting from different sequences of pulses applied to 
	the thermal equilibrium density matrix 
	$\rho_{thermal}=I_z^1+I_z^2+I_z^3$.  All sequences are read from
	left to right.  The reference spectra (resulting from a $\frac{\pi}{2}$ 
	pulse applied to all 3 carbons) is also shown (grey) for reference.
	Although the chemical shift is order $\frac{1}{\tau}$, no 
	significant phase evolution is seen.  Selective coupling 
	sequences are also demonstrated.}
   \label{Spectra}
\end{figure}

First, a series of selective single-spin pulses were applied to the
thermal state to demonstrate that these selective rotations fully
refocus the internal Hamiltonian and so can be concatenated
arbitrarily.  Fig.~\ref{Spectra} shows sample sequences and the
resulting spectra.  In addition, because the internal Hamiltonian is
fully refocused, applying selective transformations on different spins
sequentially has an effect equivalent to applying all the appropriate
transformations to each of the spins simultaneously (neglecting
relaxation).  With simultaneous, fully self-refocused $\pi$ pulses,
selective couplings can be efficiently implemented using previously
published techniques~\cite{ErnstBook,Jones}, and is demonstrated in 
Fig.~\ref{Spectra}.

Second, the projection and attenuated correlation between the expected and
experimentally determined results of different control sequences were
measured for a set of input states.  While spectra contain information
about the observables of the current state of the system
(single-spin transitions in the case of NMR), a single spectrum does
not contain enough information to reconstruct the entire state of the
system (density matrix).  By using ``read-out'' pulses that rotate
different elements of the density matrix into the single-spin
transitions, every term of the density matrix can be
determined~\cite{Tomography}.  In our case, seven repetitions of the
experiment, each with a different readout pulse, are sufficient to
reconstruct the density matrix~\cite{ReadOutPulses}.  Using
this method to determine $\rho_{out}$, the projections and attenuated 
correlations, averaged over three different input states, and under the action
of six different pulses, are measured and listed in Table~\ref{ExpResults}.  
The input states considered are of the form 
\beqn
\rho_{in}=I_j^1+I_j^2+I_j^3, 
\eeqn 
for $j={x,y,z}$.

Because the pulses are short with respect to the decoherence times, 
which are all greater than 200ms, the difference between the projection 
and the correlation indicates incoherent evolution (e.g., RF inhomogeneity) 
has caused significant errors.  This is demonstrated by the fact that the 
ratio of the correlation to the projection, or the attenuation, is on the 
order of the projection (see Table~\ref{ExpResults}).  This further supports 
the need to address errors caused by inhomogeneous effects.  Robustness to 
inhomogeneity can be added as a criterion in pulse determination; the additional 
resource requirements (time, power, etc.), however, are not known.  Decreases in 
the coherent errors are also desirable, but such improvements must first 
be sought through improved experimental implementation.  While the goal 
of fault tolerance clearly hasn't been met, these results indicate that 
experimental implementation is nearing ideal simulation results.

\section{Conclusions}
The ability to implement faithfully a desired unitary transformation
is at the heart of any future implementation of a quantum computer.
Any control technique should minimize the effects of decoherent errors
while retaining the required addressability.  We have demonstrated a
method to find control sequences that use detailed knowledge of the
system's Hamiltonian and high-power pulses to create a desired gate by 
strongly modulating the system's dynamics.  These gates are
short in duration yet selective, implementing the correct unitary
transformation.  The effect of these control sequences has been
simulated under various conditions and experimentally demonstrated by NMR.
Finally, avenues for future improvements are within the reach of
current technologies.

\section{Acknowledgements}

We are grateful to L. Viola, E. Knill and R. Laflamme for stimulating discussions.  
This work was supported by the National Security Agency and Advanced Research and 
Development Activity under Army Research Office contract number DAAD19-01-1-0519, by
the Defense Sciences Office of the DARPA under contract
number MDA972-01-1-0003, and by NSF.

\newpage
\begin{table}
\caption{Summary of the relevant characteristics for an example set of 
transformations.  The three columns list the pulse duration (in $\mu s$),
maximum power (in kHz), and the gate fidelity of the simulated pulse.
While the maximum power is relatively large, all powers are experimentally 
feasible.  The pulses designed for the Crotonic acid
sample require longer times and yield lower fidelities due to the 
decreased chemical separation and increase of coupling strengths.}
\begin{tabular}{cccc}
Pulse & Pulse Time($ \mu s $) & Max. Power($kHz$) & Fidelity \\
\hline
\multicolumn{4}{c}{\bf Alanine Pulses} \\
\vspace{.1cm} $\left. \frac{\pi}{2} \right]_x^1$ & 202& 7.9 & 0.9995 \\
\vspace{.1cm} $\left. \frac{\pi}{2} \right]_x^2$ & 221& 9.3 & 0.9995 \\
\vspace{.1cm} $\left. \frac{\pi}{2} \right]_x^3$ & 212& 9.0 & 0.9995 \\
\vspace{.1cm} $\left. \frac{\pi}{2} \right]_x^{12}$ & 194& 8.5 & 0.9995 \\
\vspace{.1cm} $\left. \frac{\pi}{2} \right]_x^{23}$ & 179& 9.2 & 0.9995 \\
\vspace{.1cm} $\left. \frac{\pi}{2} \right]_x^{123}$ & 163& 10.3 & 0.9995 \\
\vspace{.1cm} $\left. {\pi} \right]_x^{12}$ &  252& 8.0 & 0.9996 \\
\vspace{.1cm} $\left. {\pi} \right]_x^{23}$ &  129& 10.3 & 0.9997 \\
\hline
\multicolumn{4}{c}{\bf Crotonic Acid Pulses} \\
\vspace{.1cm} $\left. \frac{\pi}{2} \right]_x^1$ &  389& 8.5 & 0.9930 \\
\vspace{.1cm} $\left. \frac{\pi}{2} \right]_x^2$ &  610& 4.8 & 0.9957 \\
\vspace{.1cm} $\left. \frac{\pi}{2} \right]_x^3$ &  392& 6.3 & 0.9950 \\
\vspace{.1cm} $\left. \frac{\pi}{2} \right]_x^4$ &  559& 7.6 & 0.9923 \\
\vspace{.1cm} $\left. {\pi} \right]_x^{12}$ &  326& 9.0 & 0.9963 \\
\vspace{.1cm} $\left. {\pi} \right]_x^{23}$ &  315& 11.3 & 0.9932 \\
\vspace{.1cm} $\left. {\pi} \right]_x^{34}$ &  345& 8.4 & 0.9962 \\
\end{tabular}
\label{PulseChar}
\end{table}

\onecolumn
\begin{table}
\caption{Summary of experimental data for the Alanine sample.  
For each of six $\pi/2$ pulses, the experimentally determined 
density matrix is compared with the expected result.  
The first column (No Pulse) confirms that the experimental and 
expected inputs have almost unit overlap.  Each of the other headings
denote which spins are rotated by $\pi/2$.
In each case, the projection and attenuated correlation for each 
pulse is averaged over the three inputs 
$I_j^1+I_j^2+I_j^3$, for $j={x,y,z}$.
Because all pulses are short in comparison to the natural decoherence
times, the attenuation gives an indication of the relative significance 
of the coherent and incoherent errors.  
Statistical uncertainties arising from errors in the tomographic density 
matrix reconstruction are order 1\%.  
}
\begin{tabular}{cccccccc}
 & No Pulse & Carbon 1 & Carbon 2 & Carbon 3 & Carbons 1\&2 & Carbons 2\&3 
	& All Carbons \\
\hline
Projection	& 0.996 & 0.992 & 0.995 & 0.996 & 0.993 & 0.996 & 0.992 \\
Correlation	& NA 	& 0.977 & 0.985 & 0.986 & 0.975 & 0.984 & 0.986 \\
Attenuation	& NA 	& 0.985 & 0.990 & 0.990 & 0.982 & 0.988 & 0.994 \\
\end{tabular}
\label{ExpResults}
\end{table}

\newpage
\begin{thebibliography}{99}
\bibitem{Rabi}
	I.I. Rabi, S. Millman, P. Kusch, and J.R. Zacharias,
	{\it Phys. Rev.} {\bf 55} 526 (1939).
\bibitem{Waugh}
	U. Haeberlen, and J.S. Waugh,
	{\it Phys. Rev.} {\bf 175}, 453-67 (1968).
\bibitem{HaughtControl}
	A.F. Haught, {\it Annu. Rev. Phys. Chem.} {\bf 19}, 343 (1968).
\bibitem{WarrenControl} 
	W.S. Warren, H. Rabitz, and M. Dahleh, 
	{\it Science} {\bf 259}, 1581-1589 (1993).
\bibitem{WarrenReview}
	W.S. Warren, {\it Science} {\bf 242}, 878-84 (1988).
\bibitem{Sodickson}
	A. Sodickson, D. G. Cory, 
	{\it Prog. Nucl. Magn. Res. Spectrosc.} {\bf 33}, 77 (1998).
\bibitem{Counsell}
	C. Counsell, M.H. Levitt, and R.R. Ernst,
	{\it J. Magn. Reson.} {\bf 63}, 133-41 (1985).
\bibitem{Conolly}
	S. Connolly, D. Nishimura, and A. Macovski,
	{\it IEEE Transactions on Medical Imaging} {\bf MI-5},
	106-15 (1986).
\bibitem{Tycko}
	R. Tycko, E. Schneider, and A. Pines,
	{\it J. Chem. Phys.} {\bf 81}, 680-688 (1984)
\bibitem{Morris}
	P.G. Morris, D.E. Rourke, D.J.O. McIntyre, and A. Al-Beshr
	{\it Magnetic Resonance Materials in Physics, Biology, \& Medicine}
	{\bf 2}, 279-83 (1994).
\bibitem{Sorensen}
	J. Stoustrup, O. Schedletzky, S.J. Glasser, C. Griesinger, 
	N.C. Nielsen, and O.W. S{\o}rensen,
	{\it Phys. Rev. Lett.} {\bf 74} 2921-4 (1995).
\bibitem{Emsley}
	L. Emsley, and G. Bodenhausen,
	{\it J. Magn. Reson.} {\bf 97}, 135-48 (1992).
\bibitem{Green}
	H. Green, and R. Freeman,
	{\it J. Magn. Reson.} {\bf 93}, 93-141 (1991).
\bibitem{Ewing}
	B. Ewing, S.J. Glasser, and G.P. Drobny,
	{\it J. Magn. Reson.} {\bf 98} 381-7 (1992).
\bibitem{Schiano}
	J.L. Schiano, A.G. Webb, and R.L. Magin,
	{\it IEEE Trans. on Med. Imag.} 
	{\bf 11}, 203-14 (1992).
\bibitem{Peirce}
	A.P. Peirce, M. Dahleh, and H. Rabitz,
	{\it Phys. Rev. A} {\bf 37} 4950 (1988).
\bibitem{Zewail}
	W.S. Warren and A.H. Zewail,
	{\it J. Chem. Phys.} {\bf 78} 2279-97 (1983).
\bibitem{Dahleh}
	M. Dahleh, A.P. Peirce, and H. Rabitz,
	{\it Phys. Rev. A} {\bf 42}, 1065-79 (1990).
\bibitem{Assion} 
	A. Assion, T. Baumert, M. Bergt, T. Brixner, B. Kiefer, 
	V. Seyfried, M. Strehle, and G. Gerber, 
	{\it Science} {\bf 282}, 919 (1998)
\bibitem{Pearson} 
	B.J. Pearson, J.L. White, T.C. Weinacht, and P.H. Bucksbaum,
	{\it Phys. Rev. A} {\bf 63} 063412 (2001) 
\bibitem{Molmer}
	A. S{\o}rensen and K. Molmer,
	{\it Fortschritte der Physik-Progress of Physics} {\bf 48} 811-21 (2000).
\bibitem{Wineland}
	D. Kielpinski, V. Meyer, M.A. Rowe, C.A. Sackett, W.M. Itno, C. Monroe, and D.J. Wineland,
	{\it Science} {\bf 291} 1013-1015 (2001).
\bibitem{CoryPP} 
	D.G. Cory, A.F. Fahmy and T.F. Havel, 
	{\it Proc. Natl. Acad. Sci.} {\bf 94}, 1634-1639 (1997).
\bibitem{ChuangPP}
	N.A. Gershenfeld and I.L. Chuang,
	{\it Science} {\bf 275}, 350-356 (1997).
\bibitem{DJ}
	J.A. Jones, and M. Mosca,
	{\it J. Chem. Phys.} {\bf 109}, 1648-1653 (1998)
\bibitem{Grover}
	I.L. Chuang, N. Gershenfeld, M. Kubinec, and D. Leung,
	{\it Phys. Rev. Lett.} {\bf 80}, 3408-11 (1998).
\bibitem{QFT}
	Y.S. Weinstein, M.A. Pravia, E.M. Fortunato, S. Lloyd, and D.G. Cory,
	{\it Phys. Rev. Lett.} {\bf 86} 1889-91 (2001). 
\bibitem{3BitQEC} D. G. Cory, W. Maas, M. Price, E. Knill, R. Laflamme, 
	W. H. Zurek, T. F. Havel and S. S. Somaroo, {\it Phs. Rev. Lett.}
	{\bf 81}, 2152-2155 (1998)
\bibitem{5BitQEC}
	E. Knill, R. Laflamme, R. Martinez, and C. Negrevergne,
	{\it Phys. Rev. Lett.} {\bf 86},  5811  (2001).
\bibitem{QSim}
	S. Somaroo, C.-H. Tseng, T.F. Havel, R. Laflamme, and D.G. Cory,
	{\it Phys. Rev. Lett.} {\bf 82}, 5381-84 (1999).
\bibitem{Bloch-Siegart}
	F. Bloch and A. Siegert,
	{\it Phys. Rev.} {\bf 57}, 522-527 (1940).
\bibitem{ChuangPulses}
	M. Steffen, L.M.K Vandersypen, and I.L. Chuang,
	{\it J. Magn. Reson.} {\bf 146}, 369-74 (2000).
\bibitem{KnillPulses}
	E. Knill, R. Laflamme, R. Martinez and C-H. Tseng,
	{\it Nature} {\bf 404}, 368-70 (2000).
\bibitem{PulseProblem}
	For instance, not all errors can be represented as a composition 
	of phase shifts, $\sigma_z\sigma_z$ couplings and ideal 
	$\pi/2$ or $\pi$ pulses.
\bibitem{Hahn}
	E.L. Hahn,
	{\it Phys Rev.} {\bf 9080}, 580-594 (1950).
\bibitem{CP}
	H.Y. Carr, and E.M. Purcell,
	{\it Phys. Rev.} {\bf 94}, 630-638 (1954).
\bibitem{MG}
	S. Meiboom, and D. Gill,
	{\it Rev. Sci. Instrum.} {\bf 29}, 688-691 (1958).
\bibitem{Eraser}
	G. Teklemariam, E.M. Fortunato, M.A. Pravia, T.F. Havel
	and D.G. Cory,
	{\it Phys. Rev. Lett.} {\bf 86}, 5845-49 (2001).
\bibitem{NS}
	L. Viola, E.M. Fortunato, M.A. Pravia, E. Knill, R. Laflamme, and D.G. Cory,
	{\it Science} {\bf 293} 2059 (2001).
\bibitem{Spect}
	J. Huth, N.D. Kurur, and G. Bodenhausen,
	{\it J. Magn. Reson.} {\bf 118}, 286-90 (1996).
\bibitem{image}
	R.E. Gordon, P.E. Hanley, D. Shaw, D.G. Gadian, G.K. Radda, 
	P. Styles, P.J. Bore, and L. Chan,
	{\it Nature} {\bf 287} 736 (1980).
\bibitem{Universality}
	S. Lloyd,
	{\it Phys. Rev. Lett.} {\bf 75}, 346-9 (1995).
\bibitem{Schumacher}
	B. Schumacher,
	{\it Phys. Rev. A} {\bf 54}, 2614-28 (1996).
\bibitem{PowerNote}
	Actually, $\omega$ equals a spin's nutation rate caused by an RF field.  
	Because this parameter is experimentally controlled by attenuating 
	the RF power, it is commonly referred to as the pulse power.
\bibitem{Simplex}
	J.A. Nelder and R. Mead, 
	{\it Comput. J.} {\bf 7}, 308-13 (1965).
\bibitem{ErnstBook}
	R.R. Ernst, G. Bodenhausen, and A. Wokaun,
	{\it Principles of Nuclear Magnetic Resonance in One and Two Dimensions.}
	Oxford University Press, Oxford (1994).
\bibitem{Fault-Tolerance}
	E. Knill, R. Laflamme, and W.H. Zurek,
	{\it Science} {\bf 279}, 342-5 (1998).
\bibitem{CoryRFSelection}
	D.G. Cory, {\it J. Magn. Reson.} {\bf 103}, 23-6 (1993).
\bibitem{Jones}
	J.A. Jones, and E. Knill,
	{\it J. Magn. Reson.} {\bf 141}, 322-5 (1999).
\bibitem{Tomography}
	I.L. Chuang, N. Gershenfeld, M. Kubinec, and D. Leung,
	{\it Proc. R. Soc. London A} {\bf 454}, 447 (1998).
\bibitem{ReadOutPulses}
	The seven readout pulses used for density matrix reconstruction are:
	$\left. \pi/2 \right]_y^1$, 
	$\left. \pi/2 \right]_y^3$, 
	$\left. \pi/2 \right]_x^3$, 
	$\left. \pi/2 \right]_y^{1,2}$, 
	$\left. \pi/2 \right]_x^{2,3}$, 
	$\left. \pi/2 \right]_y^{1,2,3}$, 
	$\left. \pi/2 \right]_x^{1,2,3}$.
\end{thebibliography}
\end{document}